\documentclass[preprint2]{aastex}

\usepackage{amssymb, amsmath, natbib, graphicx, multirow}
\usepackage{xspace}
\usepackage{relsize}
\usepackage{isotope}

\newcommand{\propyne}{CH$_3$C$_2$H}

\newcommand{\diacetylene}{C$_4$H$_2$}
\newcommand{\triacetylene}{C$_6$H$_2$}

\shorttitle{Circumstellar Chemistry of SMP LMC 11}
\shortauthors{Malek, Cami \& Bernard-Salas}

\begin{document}
\title{The Rich Circumstellar Chemistry of SMP LMC 11}

\author{S.E.~Malek}
\affil{Department of Physics \& Astronomy, University of Western
  Ontario, London, ON N6A 3K7, Canada}
\email{smalek2@uwo.ca}

\author{J.~Cami} 
\affil{Department of Physics \& Astronomy, University
  of Western Ontario, London, ON N6A 3K7, Canada} 
\affil{SETI
  Institute, 189 Bernardo Ave, Suite 100, Mountain View, CA, 94043,
  USA} 
\email{jcami@uwo.ca} \and \author{J.~Bernard-Salas}
\affil{Institut d'Astrophysique Spatiale, CNRS/Universite Paris-Sud
  11, 91405 Orsay, France}

\begin{abstract}
 Carbon-rich evolved stars from the asymptotic giant branch to the
 planetary nebula phase are characterized by a rich and complex carbon
 chemistry in their circumstellar envelopes.  A peculiar object is the
 preplanetary nebula SMP LMC 11, whose Spitzer-IRS spectrum shows
 remarkable and diverse molecular absorption bands. To study how the
 molecular composition in this object compares to our current
 understanding of circumstellar carbon chemistry, we modeled this
 molecular absorption.  We find high abundances for a number of
 molecules, perhaps most notably benzene.  We also confirm the
 presence of propyne (CH$_3$C$_2$H) in this spectrum.  Of all the
 cyanopolyynes, only HC$_3$N is evident; we can detect at best a
 marginal presence of HCN.  From comparisons to various chemical
 models, we can conclude that SMP LMC 11 must have an unusual
 circumstellar environment (a torus rather than an outflow).
\end{abstract}


\section{Introduction}

As stars with initial masses between 0.8 and 8--9 M$_{\sun}$ approach
the end of their lives, they reach the asymptotic
giant branch (AGB) stage.  The AGB is characterized by alternate
hydrogen and helium shell burning, dredge-up events and extreme 
mass-loss rates \citep[up to 10$^{-4}$ M$_{\sun}$ year$^{-1}$,][]{Ibe83}.
This high mass loss causes the star to evolve further,
leaving the AGB.  When a star leaves the AGB it may become what is
known as a preplanetary nebula (pPN) during a relatively
short-lived \citep[lasting $\sim$10$^3$--10$^4$ years,][]{VW94}
transitionary period before becoming a planetary nebula (PN).

The material lost by a star during its time on the AGB goes into the
circumstellar environment (CSE) before it is dispersed into the
interstellar medium (ISM).  The CSE is a relatively cool region (with
temperatures lower than the effective temperatures of AGB stars, where
T$_{\rm eff} \approx$ 3000 K), which allows the formation of molecules
and dust beginning on the AGB and continuing into the (p)PN stages.

One of the first and most stable molecules to form in the CSE is CO.
As a result, the relative amounts of carbon and oxygen in the CSE
largely determines future chemistry.  Stars begin their lives with
more oxygen than carbon (C/O $<$ 1), but depending on the initial
stellar mass and metallicity, an AGB star may undergo sufficient dredge-up
events and become carbon-rich (C/O $>$ 1), resulting in what is known as a
carbon star (or a carbon-rich star).

The chemistry of carbon stars can result in an assortment of
molecules due to the ability of carbon to form a variety of chemical
bonds.  For example, more than 60 molecules have been detected in the
CSE of the prototypical carbon-rich AGB star IRC+10216
\citep[e.g.][]{Che96}.

The carbon chemistry in the CSE of AGB stars and (p)PNe is also of
particular interest because a group of molecules known as polycyclic
aromatic hydrocarbons (PAHs) are thought to form here \citep{Lat91}.
These molecules are ubiquitous in the universe with up to 10--15\% of
cosmic carbon contained in PAHs \citep{SW95}.

In the current models of PAH formation, benzene (C$_6$H$_6$) formation is
considered a bottleneck step \citep{Fre89, Che92}.

Not much is known about the formation of benzene in the CSEs of
evolved stars, although there are chemical models describing its
formation in the inner shocked regions of the CSE surrounding AGB
stars \citep{Che92} as well as through photochemical reactions in the
CSE of pPNe \citep{Woo02, Woo03}.

While PAHs are ubiquitous in the universe (including carbon-rich PNe),
benzene is not often found in evolved stars; it has been found in just
two objects thus far: \object{CRL 618} \citep{Cer01a} and \object{SMP
  LMC 11} \citep[hereafter Paper I]{BS06}. This suggests that it is 
either difficult to form or that it reacts quickly once formed.

In this paper, we will discuss the latter object, SMP LMC 11, which is
a carbon-rich pPN in the Large Magellanic Cloud. It is described as a
low-excitation pPN \citep{SMP78, Mor84} and has a bipolar outflow 
\citep{Sha06}. It also shows a rather high expansion velocity 
\citep[122 km/s, ][]{Dop88} with multiple velocity components. 

Here we present a detailed analysis of the molecular absorption bands
in the mid-infrared spectrum of SMP LMC 11 (first presented in Paper I) 
and show that our results are inconsistent with current models for the 
chemistry in evolved carbon-rich CSEs.

We begin this paper with a description of the
observations and data reduction in Section \ref{sec:observations},
then we describe the dust continuum in Section \ref{sec:dust}.  We
follow this with an inventory of the molecular bands in the spectrum
and the method we use to model these bands in Section
\ref{sec:molecules}.  We then present our results from our model fits
in Section \ref{sec:results}.  Next, we discuss the implications of
our results for the evolutionary status, chemical evolution and
geometry of SMP LMC 11 in Section \ref{sec:discussion}.  Finally, we
present our conclusions in Section \ref{sec:conclusion}.

\section{Observations and Data Reduction}
\label{sec:observations}

SMP LMC 11 was observed with the Spitzer Space Telescope \citep{Wer04}
Infrared Spectrograph \citep[IRS,][]{Hou04} as part of the GTO program
on June 6, 2005 (program ID 103, AOR key 4947712).  Here we present a
new reduction of the spectrum with the latest calibration files
(pipeline version S18.18).  We obtained the basic calibrated data
(BCD) files for SMP LMC 11 and processed the data for the short high
(SH, R = 600, $\lambda$ = 9.9--19.6 $\mu$m), short low (SL, R =
60--127, $\lambda$ = 5.2--14.5 $\mu$m) and long low (LL, R = 57--126,
$\lambda$ = 14.0--38.0 $\mu$m) modes.

We cleaned the data using {\sc irsclean} with the campaign rogue pixel
mask and extracted it in {\sc smart} v8.2.1 \citep{Hid04}; we
extracted the SH data using full aperture extraction, and the SL and
LL data with the manual optimal extraction mode \citep{Leb10}.  Next
we defringed the LL mode and trimmed the edges of the orders for all
modules to remove edge effects.  We eliminated flux jumps between the
orders by comparing the overlap regions and scaling the orders (for
the SH, orders were scaled to match order 20, for the low resolution
data SL2 was scaled to SL1 and both were scaled to the LL data), then
we averaged the flux from the two nod positions.  Since we were unsure
of the reliability of the initial uncertainty estimates, we instead estimated
the uncertainties on the flux values by measuring the standard
deviation in a featureless region of the SH spectrum between 16.53 and
17.44 $\mu$m; we found a standard deviation of 0.0122 Jy, corresponding
to a S/N ratio of 47 in this range.

\begin{figure}[t!]
\includegraphics[scale=0.5]{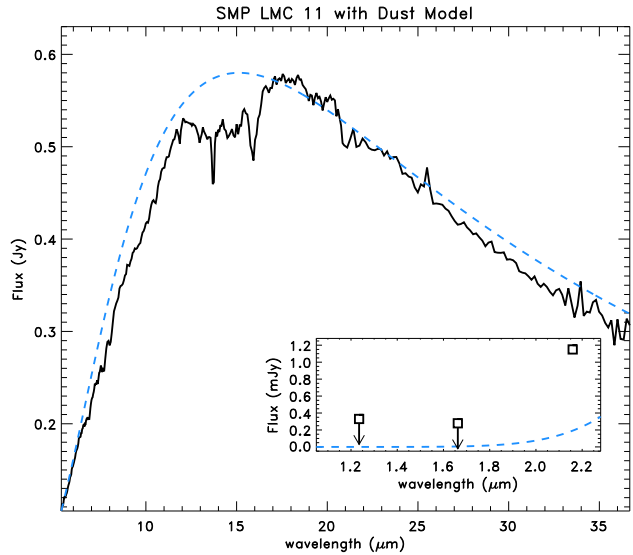}
\caption{The Spitzer-IRS low-resolution spectrum (SL and LL combined)
  of SMP LMC 11 (black) and a dust model (dashed grey or dashed
  blue line in the electronic edition). The inset shows the 
  2MASS fluxes with the dust model. See the electronic version of the
  Journal for a color figure. }
\label{fig:dust}
\end{figure}

Finally, to facilitate comparison to molecular models, we
shifted the spectrum to the rest frame using a radial velocity of
263.5 km s$^{-1}$ \citep{Mor98} and the relative motion of Spitzer at
the time of the observations (V$_{\rm LSR}$ = 12.5 km/s).  The full
low resolution spectrum is shown in Fig.~\ref{fig:dust}.
 
\section{The dust continuum}
\label{sec:dust}

Fig.~\ref{fig:dust} shows the Spitzer-IRS continuum as well as the
2MASS \citep{Skr06} H, J and K band fluxes \citep{Cut03}. From those
data, it seems clear that there is no appreciable stellar continuum
that could contribute to the mid-IR emission. We thus conclude that
the flux in the entire IRS spectrum is dominated by dust emission.

The region beyond 17 $\mu$m has a few apparent features near 19 and 
21 $\mu$m in the LL data.  However, as these features are absent from 
a quick reduction of the long high (LH, R = 600, $\lambda$ = 18.7--37.2 
$\mu$m) data, we consider these to be artifacts.  Thus, the dust emission 
does not show clear spectral features, and thus is presumably caused by 
amorphous carbon dust.  

Since it is precisely this dust emission that is then absorbed by
molecular bands (especially in the 10--17 $\mu$m region, see
Fig.~\ref{fig:dust}), we require some idea about the properties of
this dust continuum before we can model the molecular bands. A single
blackbody curve cannot properly reproduce the overall shape of the
dust continuum, suggesting that there is some stratification in the
dust layers. From our point of view, we can see through the outer (and
colder) dust layers up the point where the dust becomes optically
thick; at that point, the dust temperature is $T_{\rm d}^{\rm
  max}$. We thus approximated the dust emission by a weighted sum of
blackbody spectra at temperatures between $T_{\rm d}^{\rm max}$ and an
arbitrarily chosen minimum temperature $T_{\rm d}^{\rm min}$ of 25 K
in steps of 25 K. Assuming optically thin dust in radiative
equilibrium with the hottest (optically thick) dust, we then
determined the appropriate weights for each temperature bin consistent
with a constant-velocity outflow at a constant mass-loss rate (i.e. we
determined the mass in each layer). We find that the best model is one
with a maximum dust temperature of about 425 K (see
Fig.~\ref{fig:dust}).  However, even in such a case, we overestimate
the dust emission at the longer wavelengths; this suggests that there
is less cold dust than expected for a constant-velocity outflow at a
constant mass-loss rate. Nonetheless, the cold dust does not matter
much for the analysis of the molecular bands; for the modeling
presented in this paper, we assume that the molecules are absorbing
425 K blackbody radiation.

We also note that our dust temperature estimate is higher than that in
Paper I, but our estimate should be more reliable than their 
greybody fit.

\section{Molecular bands}
\label{sec:molecules}

\subsection{Molecular Inventory}

Superposed on the dust emission are many absorption features due to
various molecular species (see Fig.~\ref{fig:all_fits}). We will 
restrict our discussion to wavelengths shorter than
17 $\mu$m, where most of the absorption occurs.  

Here, we describe the molecular bands in great detail and determine 
the physical conditions by modeling these bands.

Acetylene (C$_2$H$_2$) is often the dominant molecular absorber in
the infrared spectra of carbon-rich pPNe. In the spectrum of SMP LMC 11,
C$_2$H$_2$ provides by far the strongest absorption. The Q-branch for
the $\nu_5$ bending mode is obvious as a very deep absorption
feature at 13.7 $\mu$m; additionally, the P- and R-branches of
C$_2$H$_2$ cause much of the broad and deep absorption that is obvious
between 12 and 16 $\mu$m (see Fig.~\ref{fig:dust}). Such broad and
deep C$_2$H$_2$ absorption is also seen in some other carbon-rich
objects, such as the carbon star \object{IRAS 04496-6958}
\citep{Spe06}, for example.  In addition, acetylene exhibits a much
weaker combination band ($\nu_4$+$\nu_5$) at 7.5 $\mu$m which is
blended with other features in the spectrum of SMP LMC 11.

Larger acetylene chains are less commonly observed in the CSE of
pPNe \citep{Fon11}. In the spectrum of SMP LMC 11 though, diacetylene 
(C$_4$H$_2$) is another major contributer to the molecular absorption in the
spectrum. The $\nu_8$ bending mode of this species appears at 15.9
$\mu$m and the $\nu_6$ + $\nu_8$ combination band is clearly visible
at $\sim$8 $\mu$m. Small amounts of the corresponding $\nu_{11}$ 
bending mode of triacetylene (C$_6$H$_2$) might be present in the red
wing of the 15.9 $\mu$m feature.

In contrast, HCN---which often shows up in carbon star spectra
alongside the C$_2$H$_2$ band at 14 $\mu$m---at best barely
contributes to the absorption in this spectrum. It is thus surprising
that HC$_3$N is present and shows a clear and deep absorption at 15.03
$\mu$m from its bending mode. There is no evidence for any longer
cyanopolyynes; for instance, there is no absorption at 15.57 $\mu$m
from the $\nu_7$ band of HC$_5$N.

Nearly unique for pPNe \citep[with one other detection thus far in CRL
618, see][]{Cer01a}, the spectrum also shows significant benzene
absorption (as noted in Paper I).  It has its deepest
absorption at 14.85 $\mu$m, where the $\nu_4$ bending mode absorbs
more than 15\% of the total continuum flux!  Additionally, strong
absorption is clearly present at the wavelengths where other strong
benzene bands are expected: another bending mode ($\nu_{14}$) at
9.6 $\mu$m and a ring stretching and deforming mode ($\nu_{13}$) at
6.72 $\mu$m.

In the blue wing of the C$_4$H$_2$ band, near 15.78 $\mu$m, some
additional absorption could be due to the $\nu_8$ bending mode of
propyne (CH$_3$C$_2$H, also sometimes called methylacetylene) at 15.78
$\mu$m, as suggested in Paper I.  Furthermore, there is also
some absorption visible in the spectrum from the $\nu_4$ band of
CH$_4$ at 7.7 $\mu$m as well as the $\nu_7$ band of C$_2$H$_4$ at
10.53 $\mu$m. Finally, we could not determine the origin of
the absorption feature near 10.38 $\mu$m, although we suspect a
molecular origin for this feature.

\subsection{Modeling the molecular absorption}

We modeled the molecular absorption using the same methods that are
used to build the SpectraFactory database \citep{Cam10}.  These model
calculations start from line lists detailing the frequencies and
intensities of the individual molecular transitions.

Line lists for C$_2$H$_2$ (including the H$^{13}$CCH
isotopologue), HCN, CH$_4$ and C$_2$H$_4$ are taken from the HITRAN
2008 database \citep{Rot09}; the line lists for C$_4$H$_2$,
C$_6$H$_6$, HC$_3$N and CH$_3$C$_2$H are from the GEISA database
\citep{Jac08}.

As we could not find reliable line lists for all species or bands, we
calculated some line lists from molecular constants and {\sc pgopher} v
7.1.108 \citep{pgopher}. The GEISA line list for benzene contains only
data for the fundamental $\nu_4$ band.  Thus, in order to model
benzene absorption at the shorter wavelengths in the SL data, we
calculated line lists for the $\nu_{13}$ and $\nu_{14}$ bands using
molecular constants from \citet{Dan89b}. Similarly, the GEISA line
list for C$_4$H$_2$ does not contain data for the transitions of the
$\nu_6+\nu_8$ combination band at $\sim$8 $\mu$m; we thus calculated a
line list using molecular constants found in \citet{AJ92,
  Gue84,Khi95}. Finally, we calculated a C$_6$H$_2$ list using data
from \citet{MB91}.

From these line lists, we calculated optical depths assuming a
population in local thermodynamic equilibrium and a Gaussian intrinsic
line profile with a width of 10 km s$^{-1}$, which is typical of
outflows of evolved stars. We carried out radiative transfer through
isothermal, plane-parallel molecular slabs in front of a 425 K
blackbody background (see Section~\ref{sec:dust}) and smoothed and
rebinned the resulting models to match the resolving power of the
observations---600 for the SH module and 90 for the SL data.

We compared the resulting models to the observations (normalized by a
cubic spline continuum) and calculated $\chi_{\nu}^2$, the reduced
$\chi^2$ statistic. However, we note that there may be some systematic
errors as well.  For instance, it is important to realize that the
current line lists for C$_2$H$_2$ do not allow us to reproduce the broad
and deep absorption in the 12--16 $\mu$m \citep[see][for a
  discussion]{Spe06}.

To find the best model, we calculated models at different temperatures
ranging from 200 to 400 K with a step size of 25 K, and similarly at
column densities between N = 10$^{15}$ and N = 10$^{19}$ cm$^{-2}$ in
steps of log~N = 0.1 and then calculated the $\chi^2_{\nu}$ value for each
model. For wavelength ranges containing absorption due to several
species, we properly treated line overlap by summing the optical depth
profiles for each contributing molecule prior to performing the
radiative transfer calculations when we fit several species
simultaneously. We thus simultaneously modeled the absorption of the 
C$_2$H$_2$ isotopologues and HCN between 12 and 14 $\mu$m and similarly also
combined \diacetylene, \triacetylene, CH$_3$C$_2$H and HC$_3$N in the
15--16.5 $\mu$m range. We fit benzene to the spectrum between 14.2 and 15
$\mu$m and C$_2$H$_4$ from 10.5 to 11 $\mu$m.  Where overlap appeared
across the modelled regions, we added the optical depth values
determined from earlier fits to the new regions (e.g. the best
fit optical depths for C$_2$H$_2$ and HCN were added to the optical
depth for the \diacetylene, \triacetylene, HC$_3$N, and CH$_3$C$_2$H
fits) and then performed the radiative transfer calculations.  We
expect that small errors may be introduced using this method, but due
to the small difference in temperature between the layers, this effect
should not be large.

Using the results from fitting the SH observations, we predicted the
absorption in the SL data for benzene, C$_2$H$_2$ and C$_4$H$_2$.  The
CH$_4$ absorption was fit by itself between 7.5 and 7.9 $\mu$m, but
the optical depth profiles for the $\nu_6$ + $\nu_8$ band of
C$_4$H$_2$ as well as the C$_2$H$_2$ band at 7.5 $\mu$m as determined
in the SH data were added to the total optical depth profile prior to
calculating the radiative transfer calculations for the CH$_4$ fits.

\section{Results}
\label{sec:results}

\begin{deluxetable}{l c c}
\tablecolumns{3}
\tablewidth{0pt}
\tablecaption{Temperatures and column densities for 
 the best model fits to our data as well as nominal 
 1$\sigma$ uncertainties. \label{table:fits}}
\tablehead{\colhead{Molecule}& \colhead{log~N} & \colhead{T (K)}} 	
\startdata
C$_2$H$_4$ & 17.30$_{-0.50}^{+0.05}$ & 350\phd$_{-50}^{+50}$ \\ \tableline 
C$_2$H$_2$ & 18.10$_{-0.05}^{+0.05}$ & \multirow{3}{*}{375\phd$^{+25}_{-50}$} \\
H$^{13}$CCH& 16.90$_{-0.10}^{+0.10}$ &\\ 
HCN	& 16.50$_{-1.50}$ 	 & 	\\ \tableline 
C$_6$H$_6$& 17.80$^{+0.40}_{-1.00}$ &	350\phd$_{-50}^{+50}$ \\ \tableline
C$_4$H$_2$& 17.10$^{+0.05}_{-0.05}$ & \multirow{3}{*}{325\phd$^{+12.5}_{-25}$} \\
\propyne& 17.00$^{+0.10}_{-0.05}$ &	\\
HC$_3$N	& 16.40$^{+0.10}_{-0.05}$ &	\\ \tableline
CH$_4$	& 17.80$^{+0.70}_{-0.20}$ & 250\phd$_{-12.5}^{+150}$ \\ 
\enddata
\end{deluxetable}

\begin{figure*}[!th]
\includegraphics[scale=0.55]{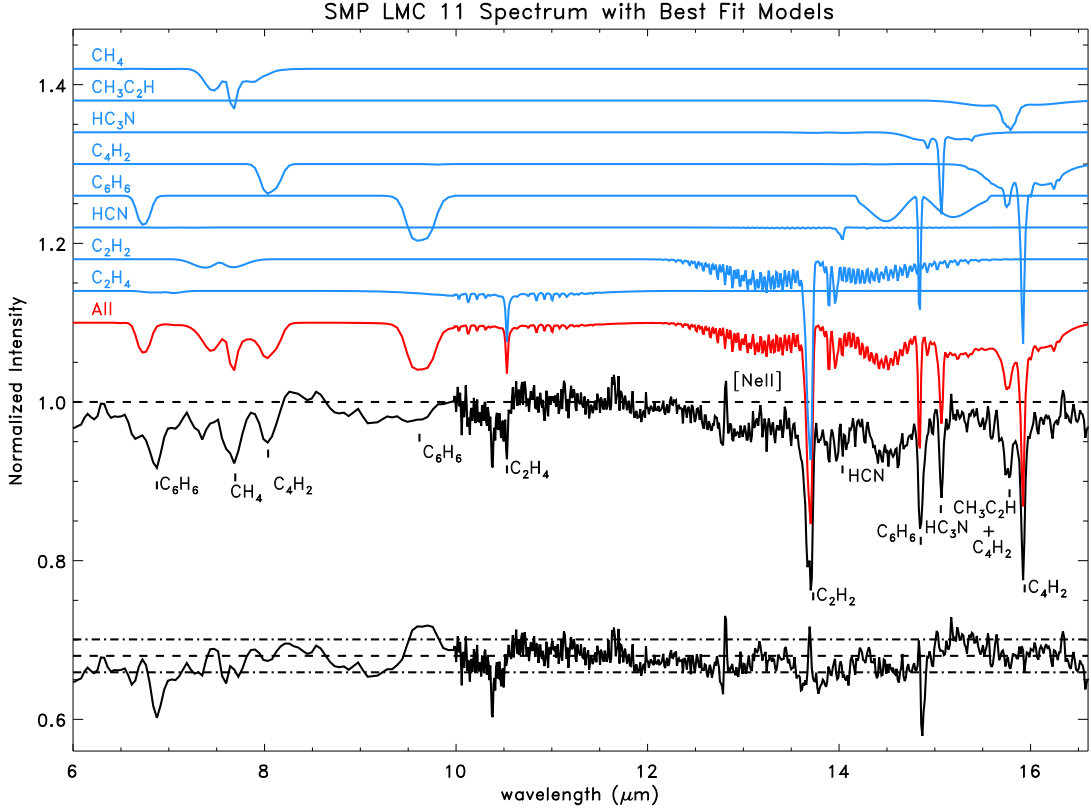}
\caption{The normalized SL and SH spectra for SMP LMC 11 from 6-16.6
  $\mu$m with our best fit models and residuals.  The combined best
  fit is shown offset above the spectrum in dark grey (red in the
  electronic edition).  The individual best fits for each molecule are
  offset above this in light grey (blue). The residuals are shown 
  offset below the spectrum and the dotted dashed lines indicate the 
  error ranges. See the electronic version of the Journal for a color 
  figure. }
\label{fig:all_fits}
\end{figure*}

\begin{figure}[t]
\includegraphics[scale=0.45]{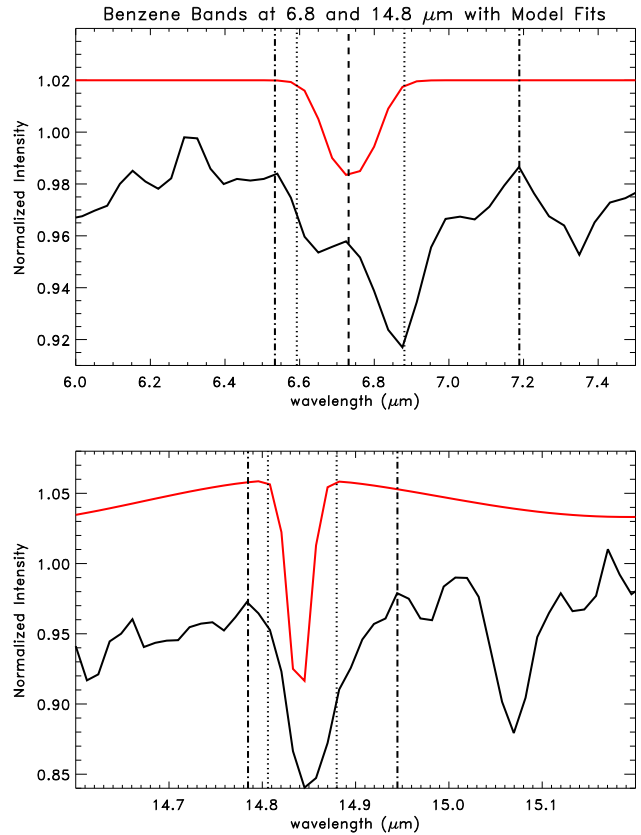}
\caption{The benzene bands at 6.8 (top) and 14.8 $\mu$m (bottom), the
  spectra are in black while the benzene models are in grey (red).
  The dotted-dashed vertical lines show the extent of the main
  absorption in the spectra, the dotted lines show the width of the
  model absorption for the entire band (at 6.8 $\mu$m) and for the
  Q-branch (at 14.8 $\mu$m).  The dashed line in the top figure shows
  the center of the absorption, coincident with the emission at the
  center of the 6.8 $\mu$m band. See the electronic version of the
  Journal for a color figure.  }
\label{fig:benz}
\end{figure}

\begin{figure}[t]
\includegraphics[scale=0.55]{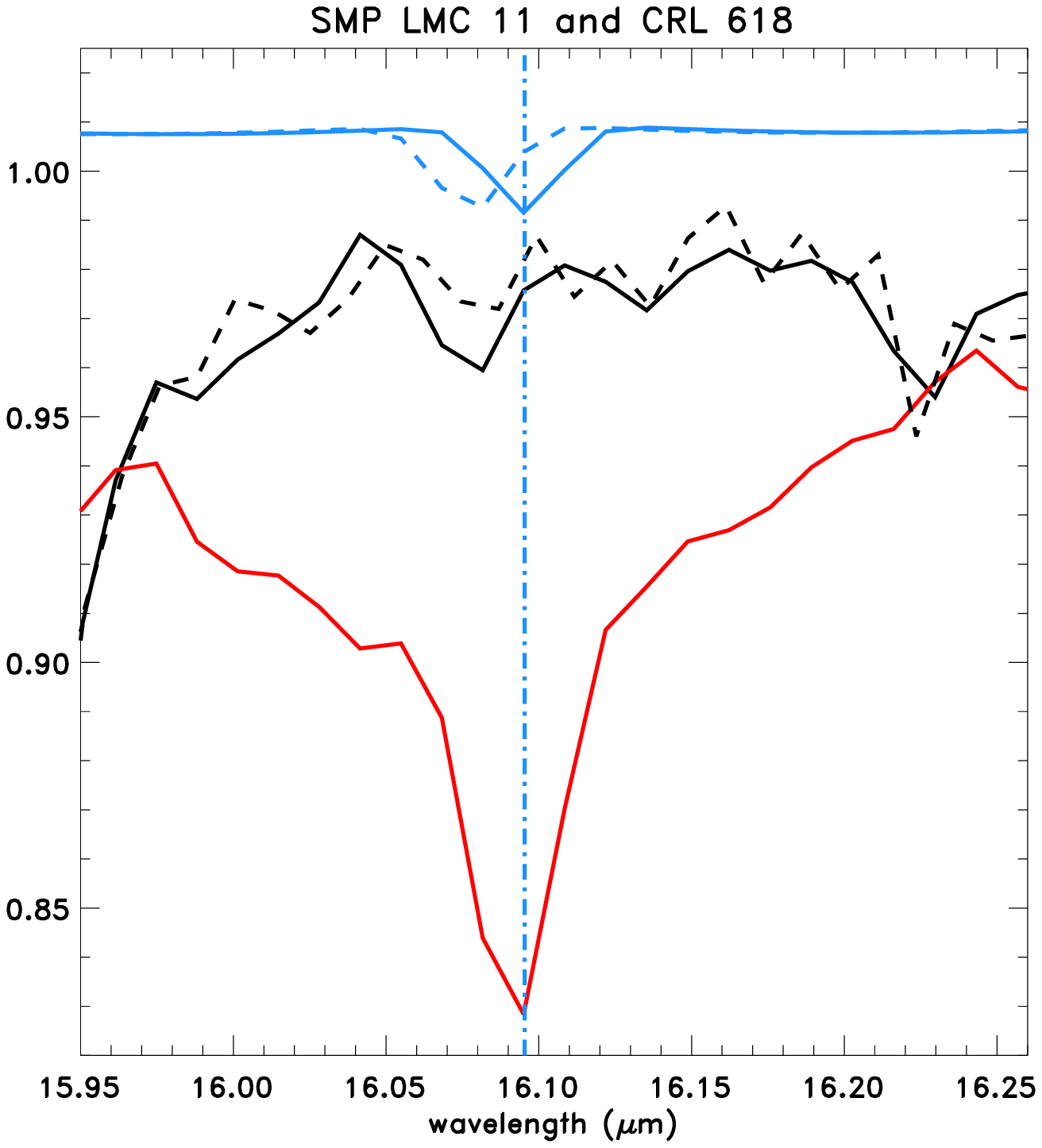}
\caption{The normalized spectra of SMP LMC 11 (black) and CRL 618 (dark grey or
  red in the electronic edition) are shown with the C$_6$H$_2$ models.
  The model calculated using molecular constants from
  \citet{MB91} is shown in the solid light grey (blue in the
  electronic edition) lines, while the modified model is in dashed
  light grey (blue) lines.  The band center for the model calculated
  with molecular constants from \citet{MB91} is highlighted with a
  light grey (blue) dotted-dashed line.  The dashed black line
  indicates an alternate fringing method for SMP LMC 11 wherein the
  appearance of the 16.07 $\mu$m feature is greatly diminished. See
  the electronic version of the Journal for a color figure. }
\label{fig:triacetylene}
\end{figure}

The best fits for the SH and SL data are shown in
Fig.~\ref{fig:all_fits}; the corresponding parameters are listed in
Table \ref{table:fits}. As can also be seen from the residuals in
Fig.~\ref{fig:all_fits}, our models reproduce the observations quite
well ($\chi^2_{\nu}$ $\approx$ 0.84 over the entire fitting
region). Our best fits indicate that the molecules are found in a
range of temperatures from 250--375 K; however, some of the
temperature stratification may be artificial due to the large
uncertainties on our temperature determinations.

\subsection{Benzene}

To fit the $\nu_4$ bending mode of benzene at 14.85 $\mu$m, we
require a high column density (log~N = 17.80) consistent with the deep
absorption in the spectrum. Our model reproduces the depth of the
absorption feature well, but does not match the width and profile
shape. This discrepancy is probably due to the absence of hot bands in our
molecular model.  Hot bands are generally slightly offset compared to
the fundamental mode, and thus tend to broaden the absorption band.
At the relatively high temperature of benzene found here, we certainly
expect some contribution from the hot bands; at 300 K, only 54\% of
the molecules should be in the ground state \citep{Kau80}. However,
hot bands are not included in the GEISA line list for benzene.  As a
result, our model fit is much narrower than it should be.

Using the benzene parameters found from fitting the $\nu_4$ band, we
predicted the appearance of the bands at 9.6 and 6.7 $\mu$m.  As seen
in Fig.~\ref{fig:all_fits}, while there are absorption features at
these wavelengths consistent with benzene, our predictions do not fit
these bands particularly well.  At 9.6 $\mu$m, we can see that the
predicted absorption is much deeper than the observed band, although
the band shape is similar.

The feature at 6.7 $\mu$m is more unusual: While there is an overall
absorption feature in the spectrum of SMP LMC 11 at this wavelength,
the feature is much broader than our model and shows what could be a
small emission bump right at the central wavelength of the benzene
absorption. Just as for the 14.85 $\mu$m band, the broader observed
feature could be a consequence of the absence of hot bands in our
model. Indeed, we note that the absorption in the spectrum appears to
be approximately twice as broad as in the model, with a similar red
degraded wing (see Fig.~\ref{fig:benz}). Thus, we consider benzene the
carrier for the overall absorption feature at 6.7 $\mu$m. The nature
of the small emission bump is not clear. If this is a real emission
feature, it seems unlikely that it would be due to benzene, since we
do not see any similar emission at the other benzene absorption bands.

\subsection{Acetylene Chains}

The C$_2$H$_2$ absorption at 13.7 $\mu$m is the deepest and broadest
absorption feature in the spectrum.  Accordingly, we find the highest 
column density for this feature (log~N = 18.10) of the molecular bands
we fit.  Since we ignored the deep and broad absorption between 12
and 16 $\mu$m which is at least partly due to C$_2$H$_2$, this is
clearly a lower limit of the true column density. In addition to the
main isotopologue, we detect a fairly high column density of
H$^{13}$CCH (log~N = 16.90) yielding a N$_{^{12}{\rm C}_2{\rm
    H}_2}$/N$_{{\rm H}^{13}{\rm CCH}}$ ratio of 16 and a
$^{12}$C/$^{13}$C ratio of 33. This is in the low end of the range for
carbon-rich objects as determined by \citet{Mil09}, who found
$^{12}$C/$^{13}$C ratios between 25-90. However, since we
underestimate the column density of the main isotopologue, this is
most certainly an absolute lower limit to the
\isotope[12]{C}/\isotope[13]{C} ratio. Similarly, from the column
densities of C$_2$H$_2$ and \diacetylene, we find a ratio of
C$_2$H$_2$/C$_4$H$_2$ of 10, which would again be a lower limit.

We find that the identification of the $\nu_{11}$ band of C$_6$H$_2$
in this spectrum is uncertain.  Using reasonable defringing
techniques, the apparent absorption feature near 16.1 $\mu$m
disappears (see Fig.~\ref{fig:triacetylene}).  Further, when the
molecular constants from \citet{MB91} are used to calculate a model
line list, we find that the calculated band center appears 0.02 $\mu$m
longer of the apparent absorption in the spectrum.  A comparison
between our calculated model and the previously identified C$_6$H$_2$
absorption in CRL 618 \citep{Cer01a} in Fig.~\ref{fig:triacetylene}
indicates that our calculated model is at least correct in peak
position, so we consider it possible that the apparent C$_6$H$_2$
absorption is not real.

If we assume that the peak near 16.07 $\mu$m is due to C$_6$H$_2$ and
shift our model of this band accordingly, the best fit is obtained at
log~N = 15.60$^{+0.70}_{-0.60}$, which does not reproduce the shape of
the band.  In order to fit the band shape, a log~N of 16.30 is
required.  However, since we have to assume that the position of our
band as well as our continuum choice is incorrect to obtain this
result, it seems unlikely that this feature is actually C$_6$H$_2$.

\subsection{Cyanopolyynes}

Although there is no obvious absorption from HCN at 14 $\mu$m, our
modeling attempts suggest that the overall fit is slightly better if
HCN is included; in fact, we find a similar (low) column density for
HCN as we do for the much stronger HC$_3$N. To verify whether the
addition of HCN {\em significantly} improves the fit, we performed an
F-test for an additional term \citep[see][]{BR03}.  We find that for
HCN, F$_\chi$ = 0.993, and the probability of achieving this value for
F$_\chi$ with the addition of a random factor in our model is 32\%.
Thus, HCN does not significantly improve the fit at this wavelength
range and the column density we find should therefore be considered an
upper limit.  Thus, the upper limit for HCN/HC$_3$N $\approx$ 1.  

We do not detect any absorption at the position of the bending mode of
HC$_5$N at 15.57 $\mu$m. Since the band strengths for the bending
modes of HC$_5$N \citep[268.2 cm$^{-2}$atm$^{-1}$,][]{Benilan2007109} 
and HC$_3$N \citep[245.1 cm$^{-2}$atm$^{-1}$,][]{Jol07} are fairly 
similar and since the modes should have a similar profile, this 
suggests that HC$_5$N is simply not present in SMP~LMC~11. From the 
measured equivalent width of the band (0.007 $\mu$m) and the 
signal-to-noise ratio of $\sim$85 in this part of the spectrum, we 
estimate an upper limit of log~N = 15.4 to the column density for 
HC$_5$N, yielding a lower limit to the HC$_3$N/HC$_5$N ratio of 10.

\subsection{Other Species}

Our best fit model shows a clear contribution
from CH$_3$C$_2$H at a relatively high column density, blended with
the C$_4$H$_2$ absorption band. Again, we performed an F-test and
found that in this case, adding CH$_3$C$_2$H does indeed significantly
improve the fit: We find F$_\chi$ = 53 and the probability of observing 
this F$_{\chi}$ value with the addition of a random factor is 
$\sim$10$^{-8}$\%. This absorption cannot be due to isotopologues of
C$_4$H$_2$ either, as these absorb at longer wavelengths than the main
isotopologue peak \citep[at 15.95 and 15.93 $\mu$m for H$^{13}$CCCCH
  and HC$^{13}$CCCH respectively, ][]{Jol10}. We thus conclude that
CH$_3$C$_2$H is indeed present in the spectrum of SMP LMC 11. Note
that this species was also observed in the pPN CRL 618 at
millimeter wavelengths, and that it was suggested to contribute to the
absorption at infrared wavelengths too \citep{Cer01b}.  Here, however,
we find a higher column density.

Finally, we also find good fits and fairly high column densities for
the CH$_4$ absorption at 7.7 $\mu$m and the C$_2$H$_4$ absorption at
10.53 $\mu$m.

Judging from the residuals and low overall $\chi^2$, we have accounted
for most of the molecular absorption.  Thus, any additional molecular
bands apart from those already noted must be either weak or perhaps
shallow and broad.

\section{Discussion}
\label{sec:discussion}

The rich molecular spectrum of SMP LMC 11 seems to offer a unique
astrophysical laboratory to study chemical pathways in carbon-rich
environments, including the formation of benzene.  Indeed, although
the CSE of SMP LMC 11 shares some properties with CRL~618---the only
other pPN in which benzene is detected---there are some
significant differences that offer clues to the conditions required
for the efficient formation of benzene and other carbonaceous
molecules.

\subsection{Circumstellar geometry}

The first aspect we consider is the geometry of the circumstellar
environment and the physical conditions of the material within. The
continuum emission in the Spitzer-IRS observations is due to dust and
provides the background intensity against which the molecular gas
absorbs. Thus, the molecular gas is either mixed in with the dust or
located further from the star.  Additionally, the dust must be
optically thick at infrared wavelengths to explain the overall shape
of the combined SL and LL spectrum.  Moreover, optically thick dust is
consistent with the featureless shape of the dust continuum which can
be represented reasonably well with blackbody curves (see also
Section \ref{sec:dust}).

It is clear that the physical conditions in the CSE of SMP LMC 11 are
somewhat different from those in CRL 618: the CSE of SMP LMC 11 is
denser and warmer than that of CRL 618. We find a typical dust
temperature of 425 K, whereas the dust temperature of CRL 618 is
98--110 K \citep{Fon11}. Similarly, the molecular excitation
temperatures are slightly higher---250--400 K versus 200--250 K---and
we also find higher column densities \citep[e.g. N(C$_2$H$_2$) = 2 
$\times$ 10$^{17}$ cm$^{-2}$ in CRL 618,][versus $\ge 10^{18}$ 
cm$^{-2}$ in SMP LMC 11]{Cer01a}.

If the CSEs of both CRL 618 and SMP LMC 11 were simple outflows, the
higher temperatures in SMP LMC 11 would indicate that the dust is
located closer to the star and thus that SMP LMC 11 turned off the AGB
more recently. Indeed, the dust and molecular gas temperatures in SMP
LMC 11 are more typical of a late AGB star than a pPN for which dust
temperatures of 150--300 K are expected \citep{Kwo00}. However, as
seen in Fig.~\ref{fig:all_fits}, the Spitzer-IRS spectrum of SMP LMC
11 also exhibits weak emission from [\ion{Ne}{2}], which is typically
one of the first excitation lines seen in young PNe. This clearly
indicates that the central star is much hotter than a typical AGB star
and is thus in the pPN stage. This apparent contradiction is easily
reconciled if we consider that the [\ion{Ne}{2}] line does not
originate from the same region as the molecular absorption and the
dust. Indeed, since the dust is optically thick, the gas from which
the [\ion{Ne}{2}] originates cannot be located in a direct line of
sight toward the central star: the dust hides the innermost region
from view and no ionizing radiation can be expected to penetrate this
thick dust layer. Thus, the [\ion{Ne}{2}] line must originate from a
region which is geometrically distinct from the location of the dust
and the molecular gas and which is not obscured by dust.

The current results suggest that the molecular gas and dust surrounding SMP LMC 
11 is located in a thick torus that we see fairly edge-on, as 
previously suggested in Paper I. In such a geometry, the gas that
causes the [\ion{Ne}{2}] emission would then be in a bipolar outflow. 
The properties of the central
object thus correspond to those of a pPN, but the circumstellar
material does not correspond to a pPN outflow and evolves on its own
timescale since it resides in a massive disk. Such a disk with
optically thick dust would also shield the circumstellar material from the
intense radiation of the central object. Note that a radially
constrained dusty torus could also explain why a dust model with a
constant-velocity outflow at a constant mass-loss rate
overestimates the flux at the longer wavelengths.

Thus, from the geometrical point of view, SMP LMC 11 is very similar
to CRL 618 for which a similar geometry involving a dense torus (in
addition to a bipolar outflow) is observationally established
\citep[see e.g.][]{1986MNRAS.223P..13B}. Such massive, vertically
extended and long-lived disks are often seen around binary systems
containing an evolved star such as the Red Rectangle \citep[see
  e.g.][]{1995ApJ...453..721J,Wat98}, for example. However, it is not
clear what is the primary cause for the higher temperatures in SMP LMC
11. This could simply indicate that the torus is closer to the central
star than for CRL 618, either because the CSE is younger, or because
it is expanding slower than that of CRL 618 \cite[which is expanding
  at a rate of 20 km s$^{-1}$, see
  e.g.][]{2000ApJ...530L.129H}. However, one could also attribute the
different temperatures to differences in the optical thickness of the
dust between the two objects. In either case, the molecular excitation
seems constrained to a fairly small range in temperatures.

\subsection{The chemistry in SMP LMC 11}

In terms of the molecular composition, the most striking difference
between SMP LMC 11 and CRL 618 is the much stronger absorption that we
find for benzene. The center of the benzene band at 14.85 $\mu$m is
about three times deeper in SMP LMC 11 than in CRL 618. It may be
somewhat surprising then that we find a column density that is 130
times larger than what was found for CRL 618 \citep{Cer01a}; however,
we believe that this difference is due to the fact that we used
more recent line lists. Additionally, not only is benzene itself much
more abundant in SMP LMC 11, it is also enhanced compared to most
other species, especially compared to acetylene. If we take the
lower limit for the column density of C$_2$H$_2$ at face value, we
find that there is about twice as much C$_2$H$_2$ as there is benzene;
for CRL 618, the ratio found between these molecules was about
40 \citep{Cer01a}.

The relative abundances of the acetylene chains might offer some hints
about the chemistry leading to efficient benzene
formation. \citet{Woo02,Woo03} found that the interstellar benzene
formation route \citep{1999ApJ...513..287M} was not efficient in
environments such as CRL 618. Instead, they proposed a much more
efficient route in such environments starting from acetylene and
HCO$^{+}$. The same model also predicts high column densities of
C$_4$H$_2$ and C$_6$H$_2$ that are similar to one another. In fact,
this matches the abundances in CRL 618 to within a factor 4--6. In the
spectrum of SMP LMC 11 however, we cannot reliably establish the
presence of C$_6$H$_2$ in the spectrum.  This suggests that there is
yet another formation route to benzene in these warm and dense
environments, or possibly some additional reactions that need to be
considered. These reactions might involve fast and efficient
ring-closing reactions that deplete C$_6$H$_2$ or an additional route
to benzene formation that starts from the abundant C$_2$H$_2$ and
C$_4$H$_2$ which then decreases the efficiency of the formation of
C$_6$H$_2$.

It is equally interesting to consider the benzene loss
reactions. Chemical models for environments comparable to those
studied here show that the main destruction route for benzene is a
further reaction with CN to form benzonitrile
\citep[C$_6$H$_5$CN,][]{Woo03}.  This molecular species has strong
absorption bands at 13.2 and 14.4 $\mu$m, which we do not observe in
the spectrum of SMP LMC 11. Although the CN in these reactions is
expected to originate from the photodissociation of HCN which is again
at best marginally present in the spectrum of SMP LMC 11. Benzene
destruction could thus be inhibited by the absence of the CN parent
species.

\bigskip

The cyanopolyynes also present an interesting case. As shown in
\citet{CG93}, for example, the primary routes to the formation of
cyanopolyynes in pPNe environments are through reactions between
members of the acetylene family and the CN radical which itself is
produced from HCN in photochemical reactions. Reactions between CN and
C$_2$H$_2$ then result in HC$_3$N, while reactions with C$_4$H$_2$
yield HC$_5$N. Since both C$_2$H$_2$ and C$_4$H$_2$ are abundant in
the spectrum of SMP LMC 11, we would thus expect fairly large
abundances of the longer cyanopolyynes as well.  However, while we
clearly detect a strong absorption band due to HC$_3$N, we do not find
much evidence for the parent molecule, HCN, nor for the longer
cyanopolyynes. HC$_5$N, for instance, should have a strong bending mode
at 15.57 \micron\ \citep{Benilan2007109} which does not appear in the
spectrum (see Fig. \ref{fig:all_fits}).

Such abundance patterns are very different from those observed in
CRL 618 and cannot be accommodated by the chemical models for these
environments \citep[e.g.][]{CG93,Woo03}. Even if one would consider
that HCN could be completely depleted by HC$_3$N formation, one
would expect to see efficient formation of HC$_5$N as well
since this involves the same mechanism and large amounts of C$_4$H$_2$ 
are available for this process. This observation then suggests that 
for the cyanopolyynes, some of the chemical pathways that are 
possible in pPNe environments might be missing from the models.

\subsection{Pathways to PAHs?}

The formation of benzene is often studied in the context of PAHs, for
which they are the basic unit. In the formation pathways for PAHs, the
formation of benzene is considered to be the bottleneck \citep[see
  e.g.][]{1989ApJS...71..733A}. We searched for any features due to
neutral naphthalene and pyrene (two of the smallest PAHs consisting of
two and three aromatic rings respectively) but did not find any
evidence for these species. However, since PAH formation cannot occur 
at the low temperatures in SMP LMC 11---typically 900--1100~K is required
\citep{Fre89,1989ApJS...71..733A,Che92}, this should not be surprising.

It is interesting to consider what might happen to the benzene as the
central star and CSE evolve further. All other things being equal, the
increasingly hotter central object would heat up the CSE. If the torus
expands (as is the case for CRL 618), then the circumstellar material
will become diluted and more transparent, which would clearly increase
the importance of photochemistry. It is not immediately clear what the
result of this increased photochemistry would be, but it is certainly
possible that the conditions created are ideal for further processing
of the circumstellar benzene into PAHs. If so, this might represent an
important PAH formation pathway, provided that the torus contains
enough mass to represent a significant carbon reservoir. Note that for
objects like the Red Rectangle, the circumbinary disks are indeed
found to be massive. However, in that particular object, PAHs are
clearly found in the bipolar outflows \citep{Wat98} and not in the
torus.

Although the molecular composition of the CSE of SMP LMC 11 is
certainly unique thus far, we do not believe that the environment
represented by this object is necessarily exceptional. CRL 618 is the
only other source where benzene is detected, and it also shows different
abundances for the polyynes and the cyanopolyynes. However, the
two environments are very similar in many of their physical
properties. Given the short expected timescales for the evolution from
the tip of the AGB to the PN phase, it is reasonable to expect the 
evolution of CSEs to be fast. The two objects could then
represent cases where either the initial conditions (of the torus, for
example) were slightly different or, alternatively, they could
represent slightly different steps in the evolution of the torus. It
would certainly be interesting to study how the physical conditions of
a (slowly expanding) dense torus change in this short evolutionary
phase and how the chemistry in this environment will evolve.

\section{Conclusions}
\label{sec:conclusion}

We presented an analysis of the rich molecular absorption in the
Spitzer-IRS spectrum of SMP LMC 11. We have compared it to chemical
models for carbon-rich pPNe and to CRL 618, the only other pPN in
which benzene is detected to date. The geometrical configuration
in both objects is fairly similar and includes a dense, warm torus of
material in which a rich molecular gas resides. However, the absolute
and relative chemical abundances of the carbonaceous species in SMP
LMC 11 do not match models and are also quite different from those in
CRL 618. In particular, benzene is very abundant, making SMP LMC 11 an
important environment to consider in the study of benzene and possibly
also in PAH formation. Current chemical models can reproduce some of the
molecular absorption, but not all of it.  In particular, the absence of
C$_6$H$_2$ and HC$_5$N seems to require additional ring-closing
reactions that deplete those species. Alternatively, new chemical
pathways have to be found for the formation of C$_4$H$_2$ and HC$_3$N
that do not result in abundant formation of C$_6$H$_2$ and HC$_5$N. We
encourage a more detailed study of the chemical pathways which could
explain these anomalies.

\acknowledgements{SEM acknowledges support from an Ontario Graduate
  Scholarship in Science and Technology (OGSST). JC is supported by
  the Spitzer Space Telescope General Observer program. This
  publication makes use of data products from the Two Micron All Sky
  Survey, which is a joint project of the University of Massachusetts
  and the Infrared Processing and Analysis Center/California Institute
  of Technology, funded by the National Aeronautics and Space
  Administration and the National Science Foundation. This research
  has made use of NASA's Astrophysics Data System Bibliographic
  Services and of the SIMBAD database, operated at CDS, Strasbourg,
  France.  We would also like to thank the anonymous referee for their 
  positive feedback and constructive comments which have improved the 
  quality of this paper.  }

\bibliographystyle{NBaa}

\end{document}